# Elastic Strain Associated with Irradiation-Induced Defects in Self-ion Irradiated Tungsten

Guanze He[1], Hongbing Yu[2*], Phani Karamched[3], Junliang Liu[3], Felix Hofmann[1†]

1) Department of Engineering Science, University of Oxford, Parks Road, Oxford, OX1 3PJ, UK

2) Canadian Nuclear Laboratories, 286 Plant Road, Chalk River, ON, K0J 1J0, Canada

3) Department of Materials, University of Oxford, Parks Road, Oxford, OX1 3PH, UK

\* *hongbing.yu@cnl.ca*   † *felix.hofmann@eng.ox.ac.uk*

**Abstract**:

Elastic interactions play an important role in controlling irradiation damage evolution, but remain largely unexplored experimentally. Using transmission electron microscopy (TEM) and high-resolution on-axis transmission Kikuchi diffraction (HR-TKD), we correlate the evolution of irradiation-induced damage structures and the associated lattice strains in self-ion irradiated pure tungsten. TEM reveals different dislocation loop structures as a function of sample thickness, suggesting that free surfaces limit the formation of extended defect structures found in thicker samples. HR-TKD strain analysis shows the formation of crystallographically-orientated long-range strain fluctuation above 0.01 dpa and a decrease of total elastic energy above 0.1 dpa.

**Introduction:**

Irradiation-induced lattice defects in crystalline materials are key to understanding in-reactor property degradation of nuclear materials. These lattice defects span a wide range of sizes, and several different types can be distinguished, such as point defects, defect clusters, dislocation loops and voids. Dislocation loops have received the most attention, because they can be readily observed by transmission electron microscopy (TEM), and are formed in different nuclear materials [1,2].

TEM studies of the evolution of dislocation loop structures under different irradiation conditions, e.g. temperature, irradiation species, and irradiation dose rate, are essential for understanding defect accumulation. They provide a pathway for predicting the evolution of irradiation-induced damage in structural materials exposed to reactor environments [3] and enable validation of irradiation damage simulations [4–6]. Previous TEM studies of irradiation-induced dislocation loops have primarily focused on microstructural features, such as loop size, density, morphology, Burgers vector, nature, distribution and elemental segregation [7–10].

Elastic strain fields due to irradiation defects remain largely unexplored. Yet elastic interactions play a critical role in controlling dislocation loop morphology, evolution and the formation of self-organised dislocation structures [11,12]. This effect is particularly important at damage doses >0.1 displacement per atom (dpa) [12]. Experimentally, there are limited options for probing elastic strain fields associated with irradiation defects, though a few explorative studies have been carried out: Longer range (µm scale) strain fields have been non-invasively probed, with depth resolution, using synchrotron X-ray micro-Laue diffraction [13–15]. Multi-reflection Bragg coherent X-ray diffraction imaging (MBCDI) has been used to characterise elastic strain fields of ion-irradiation-induced defects with higher (~30 nm) 3D spatial resolution [16–18]. In self-ion-irradiated tungsten this revealed large nano-scale fluctuations of lattice strain [17]. However, MBCDI is limited in spatial resolution, requires difficult sample preparation and synchrotron access.

Cross-correlation-based high angular resolution electron backscattered diffraction (HR-EBSD) offers a high accuracy, rapid, and non-destructive tool for measuring strain on sample surfaces [19]. In self-ion-irradiated tungsten it revealed long-range strain fluctuations at 1 dpa due to defect self-organization, but failed to probe short-range lattice distortions at lower doses ($\leq$ 0.01 dpa) due to limited spatial resolution [20]. High resolution transmission Kikuchi diffraction (HR-TKD) allows much higher spatial resolution measurements of strain fields ($\leq$ 12 nm), with sensitivity to individual dislocations [21]. Here we report the use of HR-TKD to probe the evolution of elastic strain fields caused by irradiation-induced defects from low to high doses. TEM is used to image the defect structures for a direct comparison. This prototypical study focuses on pure tungsten, the main candidate for plasma-facing fusion reactor armour [22,23].

**Experimental method:**

High purity polycrystalline Tungsten (99.99 wt%) foil (120 µm thick) was punched to 3 mm discs and annealed at 1200 °C for 24 hours in vacuum. The discs were mechanically polished, and twin-jet electropolished (0.5 wt% NaOH aqueous solution, 0 °C, 14 V) to make standard 3 mm TEM samples. These were irradiated with 2 MeV self-ions at room temperature to doses of 0.01, 0.1, and 1 dpa. The anticipated displacement damage was estimated using the Stopping and Range of Ions in Matter (SRIM) code [24] (K-P calculation, threshold energy of 68 eV [15]), and a fluence of $1.51 \times 10^{14}$ ions/cm$^2$ was implanted to produce 1 dpa damage.

Dislocation loops were imaged using two-beam bright field (BF) and weak beam dark field (WBDF) on a Jeol-2100 TEM (200 kV, LaB6 source). All images were taken under g = 002, z = [001] diffraction condition. For each sample regions with different thicknesses were imaged, and the local thickness was estimated by Electron Energy Loss Spectroscopy (EELS) on a Jeol ARM 200F microscope with a Gatan GIF Quantum 965 ER with dual EELS capability.

HR-TKD measurements were performed on a Zeiss Merlin SEM with a Bruker e-flash high-resolution EBSD and OPTIMUS™ TKD head. On-axis TKD configuration was used, i.e. the optical axis of the SEM intersects the centre of the phosphor screen, with diffraction patterns collected beneath the sample. This yields the strongest diffraction signals and minimises gnomonic distortions [25]. HR-TKD analysis focussed on [001] orientated grains using the following SEM operating parameters: 30 kV, 3 nA, 800 × 600 TKD pattern size, 5 mm working distance, 8 nm step size. Cross-correlation analysis of Kikuchi patterns was performed using the CrossCourt 4 software, with the saturated central region of the patterns excluded from analysis.

**Results and Discussion:**

TEM observations of the morphology and distribution of dislocation loops in tungsten irradiated to different doses are shown in Fig. 1. The kinematic bright field images indicate that defect clustering (~ tens of nm length scale) depends on both the irradiation dose and the thickness of the sample. In the thin regions, dislocation loops appear to be isolated spots at all dose levels. The average size and number density of these loops grows with increasing irradiation dose. In the medium thickness regions, defects remain isolated spots in the 0.01 dpa sample, while some extent of defect clustering (also known as 'rafting' [7]) is observed in the 0.1 dpa sample. This is even more pronounced in the 1 dpa sample. In the thick regions, clustering of defects is visible at all dose levels from 0.01 to 1 dpa. This clustering is more prominent in bright field images, while the dark field images (contrast inverted in Fig. 1) show isolated dot-like dislocation loops with some ordering. Compared to the kinematic bright field images, the contrast in the weak beam dark field images arises from the large lattice distortion close to the core of the dislocation loops [1], thus highlighting small dislocation loops.

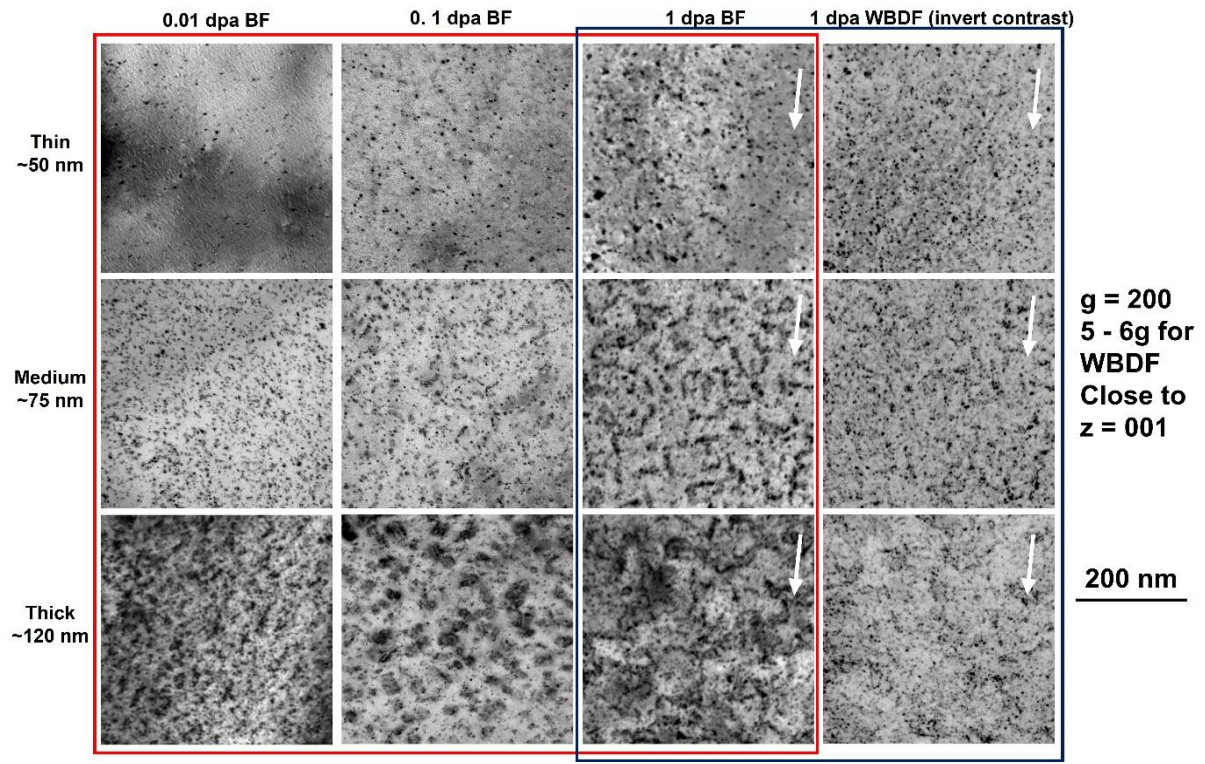

Fig. 1: TEM bright field images of the tungsten samples irradiated to 0.01, 0.1 and 1 dpa. In each sample, images were taken from thin, medium and thick regions. For the sample irradiated to 1 dpa, weak beam dark field images were also recorded at the exact same position as the bright field images.

The length scale of defect clustering visible in Fig. 1 can be quantitatively revealed by applying tangential integration to the 2D Fourier Transformation (TIFT) of the TEM bright field images (Fig. 2). For details of the TIFT method please refer to [20,26]. A characteristic length can be determined as the inverse of the spatial frequency at the peak position. For instance, the peak positions of the profiles for the 0.01 dpa, 0.1 dpa and 1 dpa sample thin regions are $1.2 \times 10^{-1}$, $7 \times 10^{-2}$, and $5 \times 10^{-2}$ 1/nm, respectively. In real space, these correspond to characteristic length scales of 8.3 nm, 14.3 nm and 20 nm. Similarly, the characteristic lengths of the defect contrast at the medium region are 20 nm, 20 nm, and 40 nm from 0.01 dpa to 1 dpa. For the thick regions, they are 33.3 nm, 55.6 nm, and 125 nm respectively. TEM image intensity fluctuation arises from two main sources, one is from the individual dislocation loops, and the other is the clustering of defects. Individual defects are small, contributing to high-frequency signals in the FFT profile, while clustering contributes to lower-frequency signals. As such, the peak in the FFT profile reflects the size of the defects' extended strain field if they are isolated (thin regions) or the scale of defect clustering if, clustering dominates

(medium and thick regions). These results suggest that the length scale of the defect contrast fluctuations increases with both dose and thickness. At the same dose level, defect clustering is more prominent in thicker samples, presumably due to a reduced effect of the free surfaces. Previous work showed that dislocation loops are attracted to free surfaces, resulting in an underestimate of dislocation loop density in TEM images [27]. The TIFT analysis shows that free surface can also limit multi-body interactions of defects and thus prevent the formation of longer-range structures. At the same thickness (except for the thinnest regions) the extent of defect clustering increases with the dose level.

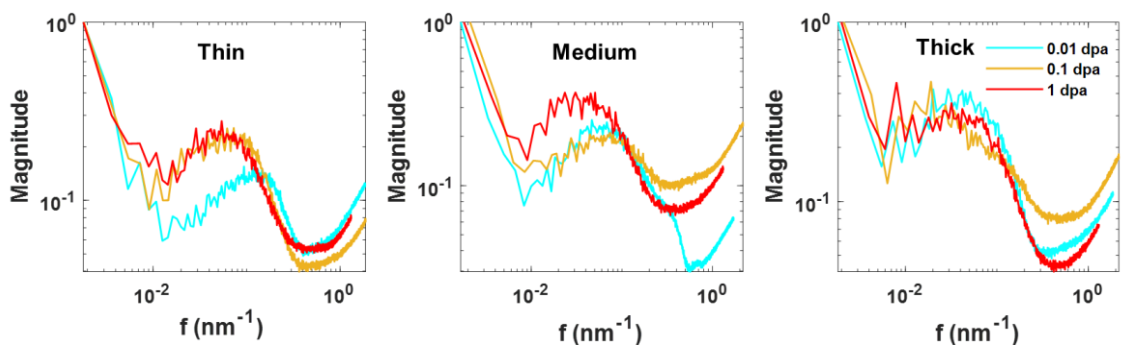

Fig. 2: Tangential integration profiles of the 2D FT of the TEM bright field images (TIFT analysis). The magnitude is normalized to 1 at zero frequency.

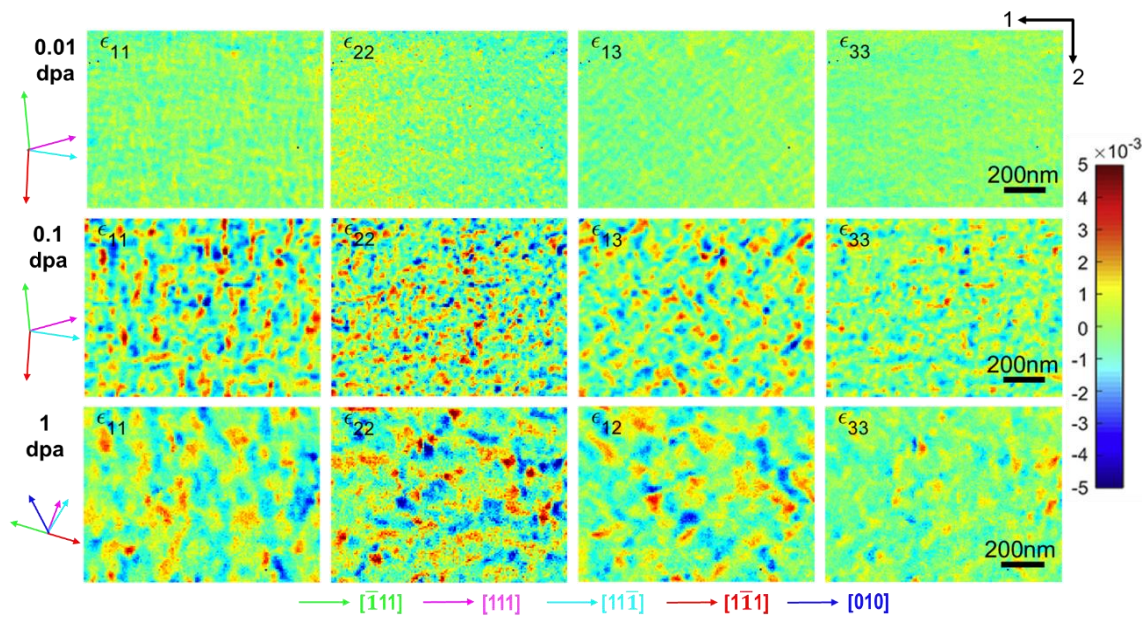

Fig. 3: Evolution of elastic strain ($\varepsilon$) as a function of irradiation dose. The coloured arrows on the left side indicate the directions of different Burgers vectors.

To explore the driving force behind the self-organization of dislocation loops, HR-TKD was used to map the 2D elastic strain fields associated with the irradiation-induced defect structures. HR-TKD was performed in the medium thickness regions, which yielded the best Kikuchi patterns. HR-TKD measures deviatoric strain, reflecting the angular distortion of lattice planes, as cross-correlation based EBSD or TKD analysis is not sensitive to lattice dilations [19]. The deviatoric strains measured in the irradiated samples are relative rather than absolute, since there is no strain-free reference point available in the irradiated samples. As such, these strain field maps reflect the spatial variation of the strain fields. Fig. 3 shows the evolution of the direct strain components and of one shear strain component with damage. All strain and rotation components are shown in the supplementary material (Fig. S1 to S3). HR-TKD data was acquired for damage levels of 0.01 dpa, 0.1 dpa and 1 dpa in grains that are approximately [001] oriented. The misorientations were respectively 22.4°, 23.7° and 5.6°. This is important as defects structures may vary with surface orientation in ion-irradiated tungsten [7].

The magnitude of elastic strains in the 0.01 dpa sample is relatively small, but they are still visible even using a colour scale of $[-5, 5] \times 10^{-3}$. The magnitude of the strain fluctuations can be quantitatively represented by the histograms of the deviation from the mean. Fig. 4 (a) shows the histograms of the deviation from the mean for the normal strain components. The full-widths-at-half-maximum (FWHMs) of the $\varepsilon_{11}$, $\varepsilon_{22}$, and $\varepsilon_{33}$ histograms for the 0.01 dpa sample are approximately $9 \times 10^{-4}$, $1.5 \times 10^{-3}$, and $7 \times 10^{-4}$, respectively. Short-range fluctuation of the lattice strain can be seen in the normal strain components and the $\varepsilon_{13}$ shear component at 0.01 dpa. TIFT analysis was used to determine the characteristic length of these short-range fluctuation (Fig. 4 (b)), which for the 0.01 dpa sample falls into range of 30 to 50 nm. This is why HR-EBSD (spatial resolution > 50 nm) failed to show strain fluctuations at 0.01 dpa [20].

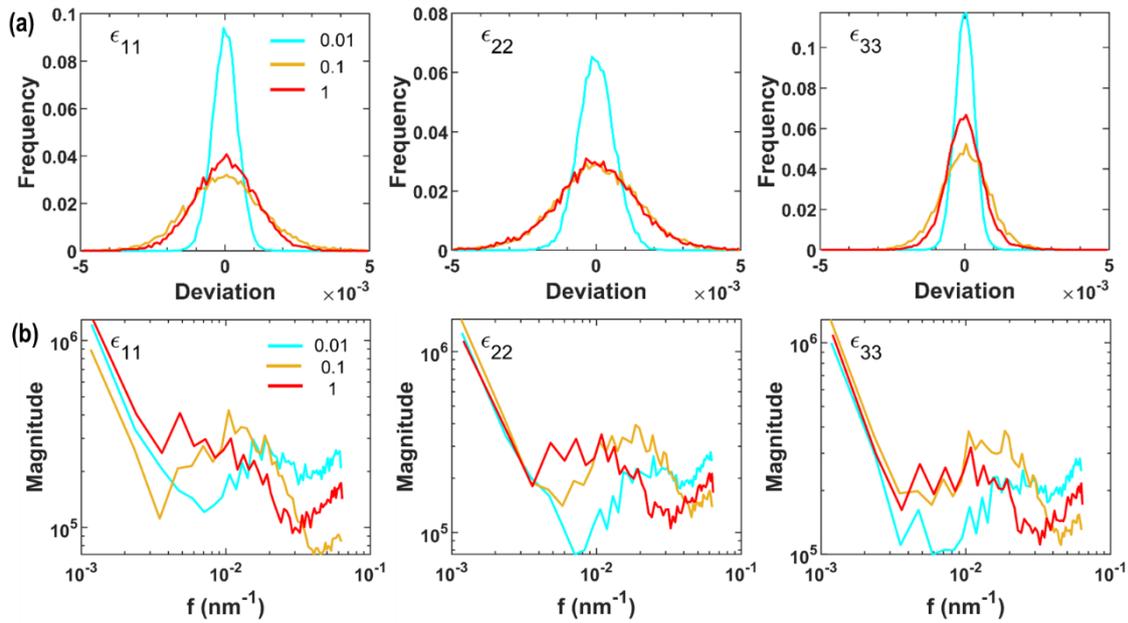

*Fig. 4: (a) Histograms of deviation from the mean for the normal strain components. (b) TIFT analysis of the normal strain maps. The colour-coding of line profiles corresponds to the different damage profiles of 0.01, 0.1 and 1 dpa.*

Strain fluctuations become more obvious in the 0.1 and 1 dpa samples (Fig. 3). This can also be seen in the histograms of the strain deviations from the mean (Fig. 4 (a)), where the FWHMs of the 0.1 dpa and 1 dpa samples are significantly greater than of the 0.01 dpa sample. With increasing dose the wavelength of the strain field fluctuations also increases, as clearly seen in the strain maps (Fig. 3) and the TIFT analysis profiles (Fig. 4 (b)). The characteristic wavelength of the normal strain fluctuations is 50 - 100 nm at 0.1 dpa and 100 – 200 nm at 1 dpa.

Interestingly, there is a clear spatial ordering of the strain fluctuations at all three damage levels (Fig. 3). At 0.01 and 0.1 dpa, the strain field of the $\varepsilon_{11}$ component predominately extends along the <111> directions, which is the energetically most favourable Burgers vector. At 1 dpa, some alignment along the <100> direction (less energetically favourable Burgers vector) is also observed, though alignment along the <111> directions still dominates.

A comparison between TEM observations and strain maps from medium thickness regions as a function of dose reveals an interesting picture: From 0.01 dpa to 0.1 dpa, there is little change in the average size of the individual dislocation loops (from 5.2 nm to 5.3 nm respectively) and the clustering size measured from the TIFT. The prominent observable change is the spatial configuration of dislocation loops, as more rafting structures are seen at 0.1 dpa (Fig. 1). However, there is a substantial increase in the magnitude of the strain fluctuations and the ordering of strain fluctuations. TKD strain measurements have a spatial resolution of ~ 12 nm [21], as such the probed strain is a convolution of the beam size with the actual strain field. For an isolated small loop, the largest strains are near the core of the loop [28]. Convolution with the electron beam will smear the strain field and give a smaller strain spread over a relatively wider range. Therefore, short-range fluctuations of smaller amplitude were observed in the strain fields of the 0.01 dpa sample – here HR-TKD effectively functions as a low-pass filter for the real strain field. The longer-range, larger amplitude strain fluctuations at 0.1 dpa are presumably due to the spatial ordering of dislocation loops driven by mutual elastic trapping [12]. The elastic trapping of dislocation loops is due to the angular dependence of elastic loop-loop interactions [5,6,11]. Langevin dynamics simulations [12] have shown that loops with collinear Burgers vectors are most strongly bound, which results in the formation of collectively stable raft structures . This is seen in the medium region of the 0.1 dpa sample (Fig. 1). These raft structures presumably result in the formation of longer range collective strain fields, over 50 nm in scale, much larger than the size of individual dislocation loops and the probe resolution.

Elastic strain reflects the elastic energy stored in the lattice. To estimate the real strain energy, measurements of the full lattice tensor are required, inaccessible to HR-TKD. Still, we can obtain the trend of the elastic energy change with damage dose by considering the sum-of-squares of the normal strains. For 0.01, 0.1 and 1 dpa we obtain 0.0107, 0.073 and 0.059 respectively. This suggests a decrease in elastic strain energy from 0.1 dpa to 1 dpa, though the wavelength of strain fluctuations increases. The maximum of elastic energy near 0.1 dpa coincides with the saturation of the mechanical properties [29] and thermal diffusivity [30] near 0.1 dpa in 20 MeV self-ion irradiated pure tungsten. However, the saturation of material properties was less clearly correlated to quantity and morphology of irradiated-induced defects, i.e. the density and size, that are normally studied. This is reasonable, especially for mechanical properties, because it is the interaction of elastic strain fields of irradiation-induced defects with glide dislocations that modify mechanical properties. The decrease in strain energy from 0.1 dpa to 1 dpa may be due to growth of interstitial loops and the formation of new lattice planes by coalescence of interstitials loops [13].

**Conclusion**

We have used TEM and HR-TKD to study the defect structures in Tungsten irradiated with 2 MeV self-ions to 0.01 dpa, 0.1 dpa and 1 dpa. Foil thickness has a profound effect on the self-organisation of irradiation defects, with free surfaces restricting the self-organisation of defects. Only thicker foils show defect rafting and formation of extended defect structures. The length scale of defect ordering increases with dose, to more than 100 nm at 1 dpa, much greater than individual cascades. HR-TKD reveals extended strain fields associated with this defect self-organisation. Their characteristic length increases with increasing dpa and is consistent with TEM observations. Surprisingly our results suggest a maximum of elastic strain energy near 0.1 dpa, coinciding with the onset of saturation of irradiation-induced property change.

**Acknowledgements**

This research was funded by Leverhulme Trust Research Project Grant RPG-2016-190 and European Research Council Starting Grant 714697. The authors acknowledge the use of characterisation facilities within the David Cockayne Centre for Electron Microscopy, Department of Materials, University of Oxford.